# Automatic detection of fiducials landmarks toward development of an application for EEG electrodes location (digitization): Occipital structured sensor based-work

E. E. Gallego Martínez, A. González-Mitjans, M. L. Bringas-Vega and P. A. Valdés-Sosa

*Abstract*—The electrophysiological source imagine reconstruction is sensitive to the head model construction, which depends on the accuracy of the anatomical landmarks locations knowns as fiducials. This work describes how to perform automatic fiducials detection, towards development of an application for automatic electrodes placement (digitization), over a three-dimensional surface of a subject head, scanned with the Occipital Inc. structure sensor ST01. We offer a wide description of the proposed algorithm to explore the three-dimensional object to features detection, by means of: dimensional reduction with perspective projection from 3D to 2D, object detection with custom detectors, robotic control of mouse motion and clicks events and re-projection from 2D to 3D to get spatial coordinates. This is done taking into account the characteristics of the scanner information, the training process of detectors with Computer-Vision Toolbox resources of MATLAB® R2018b, the integration of FieldTrip Toolbox and the main properties of several ways to represents pixels; putting together all those things to automatically find the fiducials landmarks to generate the subject's coordinate system. All this result is presented as the initial state of a project focused on developing one application for automatic electrode digitization.

The authors would like to thank for the support from the NSFC (China-Cuba-Canada) project (No. 81861128001) and the funds from National Nature and Science Foundation of China (NSFC) with funding No. 61871105, 61673090, and 81330032, and CNS Program of UESTC (No. Y0301902610100201) and the Nestle Foundation grant to MB-V and PV-S in 2018 entitled "*In Search of an EEG Neural Fingerprint of Early Malnutrition: A 50 year longitudinal Study*".

E. E. Gallego, MsC. Professor at the University of Pinar del Río ¨Hermanos Saíz¨, Telecommunications & Electronic Department, Martí final, No. 300, Pinar del Río, Cuba, CO 20100, and exchange student at the UESTC.
(e-mail: elieserernesto@gmail.com).
A. Gonzalez Mitjans is assistant professor at the University of Havana, Mathematic Department, Havana, Cuba, CO 10400, and master student of Biomedical Engineering at UESTC, Chengdu, Sichuan, China, CO 611731.
(e-mail: ani.glezmitjans@gmail.com).
M. L. Bringas-Vega. PhD. Coordinator of the Joint China-Cuba Lab for Translational Research in Neuro-technology, UESTC, Chengdu, Sichuan, China. CO 611731.
(e-mail: maria.bringas@neuroinformatics-collaboratory.org).
P. A. Valdes-Sosa. PhD. Director of the Joint China - Cuba Lab for Translational Research in Neuro-technology UESTC, Chengdu, Sichuan, China, CO 611731.
(e-mail: pedro.valdes@neuroinformatics-collaboratory.org).

*Index Terms*—Anatomical landmarks, automatic fiducials detection, computer-vision, structured sensor, EEG sensors coordinates.

## I. INTRODUCTION

An important goal for EEG –based functional brain studies is to estimate the location of brain sources produced by the scalp-recorded signals. Such source localization requires the precise locating the position of the EEG sensors; this method should be accurate, fast, reproducible, and cheap [1]. These electrode coordinates are relevant in the inverse problem solution. Inaccurate sensor co-registration (and, by extension, poor head model quality) does not only result in mislocalization of source locations, but may prevent detection of weaker signals entirely [2] although new positions represent small differences with the right one, as shown in Fig.1, in which the location of sensor A' referred to the right location in A, due to vectors magnitude overlay. Nowadays, the precise level of accuracy that is necessary or meaningful for surface electrode localization is still unclear [1].

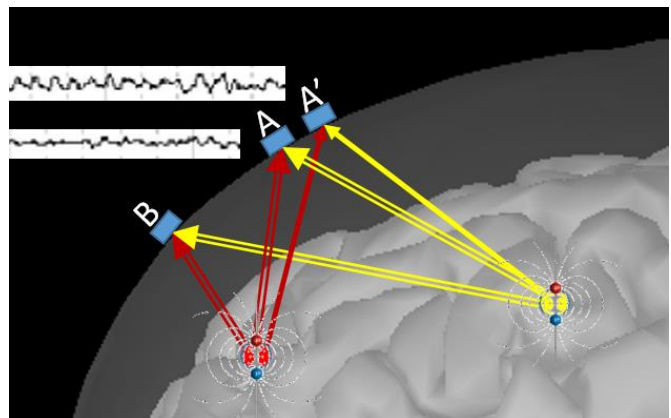

Fig.1: Effect of sources on EEG sensors, dependence on electric field overlay and electrode positions.

There are several techniques to digitize electrodes location, among them we can find manual methods, such as:

- Direct measurement [3], which consists in measuring with calipers the position between each sensor and fixed landmarks (nasion, left and right pre-auricular points) [1].
- Other method consists in measuring inter-electrode distances, although measurements are also performed using calipers, in this case the technique assumes that the EEG electrodes are positioned in a defined configuration corresponding to the 10—20 or 10—10 International System [4, 5, 6].
- Also, a free electrode placement tool is available on the FieldTrip toolbox, which allows to do a manual placement of the electrodes over the head surface, by means of a MATLAB® based script [7]. This facility has been added to the EEG-LAB toolbox as a new plugin [8]. This particular case was employed for some operations, as we will mention later.

Another alternative to this technique is classified as electromagnetic digitization, on which is based the Fastrack system. This system is based on the US Patent "Magnetic Sensor System For Fast Response, High Resolution, High Accuracy, Three-Dimensional Position Measurements" [9]. It is a 3D system, which uses a magnetic field to localize EEG electrodes. The system has a transmitter device that produces the electro-magnetic field and simultaneously constitutes a geographical reference for the positioning and orientations of the receivers. Three receivers are placed on the patient's head to carry out measurements. [1]

All the alternatives described above have in common that they need to be operated manually, which require time consuming and/or implies a very high economic cost.

Here we offer an alternative based on: The Occipital Structured Sensor ST01 for scanning the subject´s head, the Image Processing and Computer – Vision toolbox of MATLAB® for face features recognition, based on object detector, and the FieldTrip toolbox for some others particular operations.

The general idea presented here is to explore the surface of the subject's head by means of rotation, project each of the views to 2D, perform object detection to define the coordinates of the fiducials, re-project to 3D to find the coordinates spatial and with them generate the system of coordinates of the subject.

This work is the first in a series that deals with the automation of the process of digitizing the electrode coordinates in the EEG helmet. The results shown in this particular are a proof of concept as an initial state of a project aimed at the development of an application capable of digitizing the electrode coordinates automatically.

## II. MATERIALS

We used the Occipital Inc. structured sensor ST01 (Fig.2a) [10], with a precision of 0.5mm at 40cm distance [11], with an iPad A1893 to scan the subject head (Fig.2b), just as described on FieldTrip web page [12], except that we didn't add marks to the fiducials points on the subject head. Good illumination conditions and a distance between 40cm and 60cm are recommended to do the scan. Once we scan the subject, the information can be sent out by e-mail.

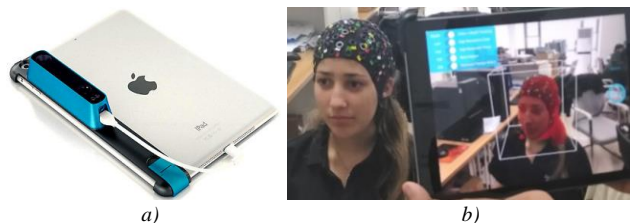

Fig.2: a) Occipital structured sensor ST01 [10] and iPad A1893 [13]. b) Scanning process around the subject.

Next section describes all performed operation in MATLAB® R2018b after scan the subject.

The experimental protocol consisted of scanning the subject's head using an EEG helmet or without it, and it was approved by the local Ethical Committee at University of Electronic Science and Technology of China, in compliance with the latest revision of the Declaration of Helsinki.

The sample consisted in nineteen healthy subjects, who were recruited voluntarily specifically for the purposes of this study. All subjects gave written informed consent for the publication of any potentially identifiable images included in this article.

We also generated a large set of images containing ears for the training of detectors, which could be available under request to the first author.

## III. METHODS

The output of the process of scanning, a compacted file is produced containing three files, all of them with the name "Model" and extensions *obj*, *jpg* and *mtl*, these files were uncompressed on the current directory of MATLAB®, and employed for the further images processing.

### A) Initialization and load the scanner information

After the MATLAB® installation contains the "*Image Processing and Computer-Vision Toolbox*", we need to initialize the FieldTrip toolbox by means of the *ft_default* function. Now, we are ready to load the scanned information to the workspace, using the function *ft_read_*headshape [7].

### B) Facial features detection: seeking the nasion.

To find the nasion and the pre-auricular points, we took into account that:

1. The nasion is the most anterior point of the frontonasal suture that joins the nasal part of the frontal bone and the nasal bones [14], and it marks the midpoint at the intersection of the frontonasal suture with the inter-nasal suture joining the nasal bones. It is visible on the face as a distinctly depressed area directly between the eyes, just superior to the bridge of the nose [15]. (See Fig.3a).

2. The pre-auricular point is at the posterior root of the zygomatic arch lying immediately in front of the upper end of the tragus [16] (Fig.3b).

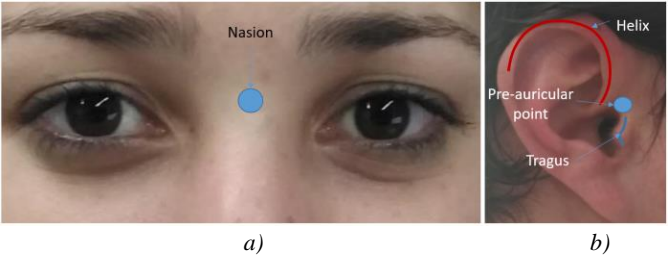

Fig.3: a) Nasion and b) Right pre-auricular point.

Following those definition we applied the following steps to locate the nasion and the pre-auricular points.

➢ To change the object orientation to display the graphical scene: it is done by the view function as *view(az,el)* setting the initial azimuth (az) to 180º, as long as the scanner software uses the iPad gyroscope while recording the scan, having as consequence that the initial view in MATLAB® is always with the head 180º down.

➢ To project the 360º rotational views of the object to frames with two dimensions and intrinsic coordinate system, using *getframe* function.

➢ To apply Computer – Vision toolbox based detector to each frame, looking for face features (eyes, nose, mouth and frontal face), or looking for "eyes pair", in both cases those features allows us to find the nasion coordinates.

For the first option we used a custom detector based on the classification models of: 'FrontalFaceCART', 'LeftEye', 'RihgtEye', 'Mouth', and 'Nose' [17, 18]. In the second option we configured a detector to use the 'EyePairBig' or the 'EyePairSmall' classification model. It was experimentally probed that the second alternative is not viable due to a very low detection rate in comparison to the first, that's why results shown on this report corresponds to the first classification model option which presents an outstanding performance as we will discuss after.

These detectors are based on Viola – Jones [20] methods, which introduces:

1. A new image representation called the "Integral Image", that allows the features used by our detector to be computed very quickly.
2. A learning algorithm, based on AdaBoost, which selects a small number of critical visual features from a larger set and yields extremely efficient classifiers [19].
3. A method for combining increasingly more complex classifiers in a "cascade" which allows background regions of the image to be quickly discarded while spending more computation on promising object-like regions. [20]

➢ After detection, we choose the best view among all the projection in which at least an object was detected, that´s why besides the eyes detector we applied the nose, mouth and face detectors, to have more decisions elements. The decision criteria to select the best view is to keep those consecutively views in which all the features were found, because of the rotation effect shown in Fig.4a), and pick the most symmetric view within the views in the optimum detection range. In some positions detectors may not see the right eye or the left eye or vice versa while they recognize some other features, depending on the sense of the rotation and the features quality in the object, as shown in the hypothetical example of Fig.4b) in detections discontinuities, and then we perform an algorithm to select a the optimum detection range among A, B and C (Fig.4b)), and finally select one view within that range views.

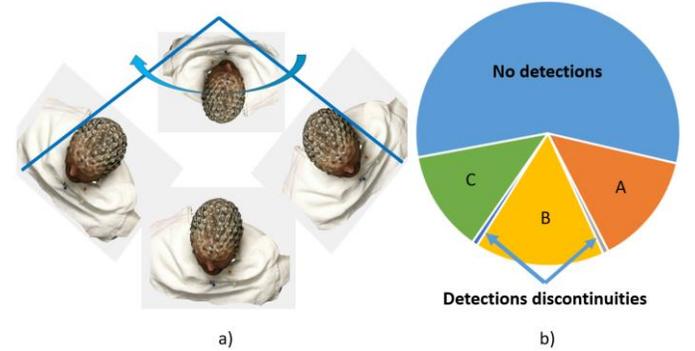

Fig.4: a) Surface rotation for features detection. b) Detection continuities and discontinuities.

Fig.5 shows the algorithm for the optimum detection arc applied once we have all the rotation angles with successful detection but with discontinuities.

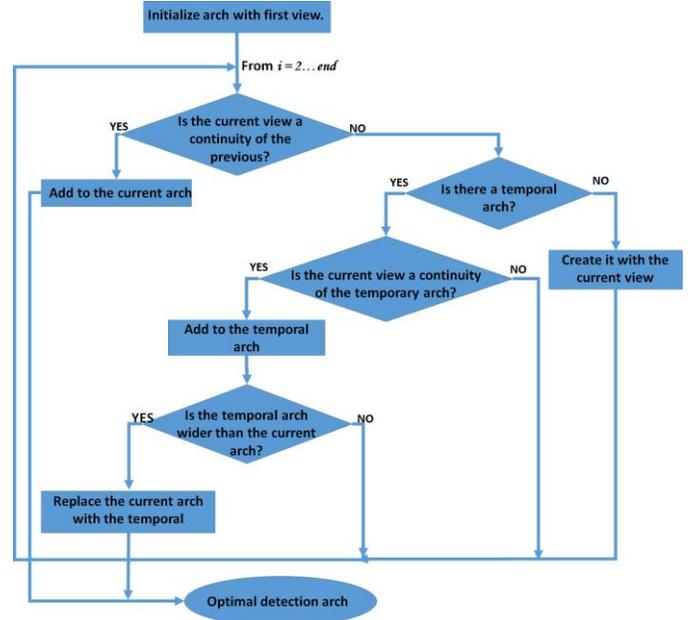

Fig.5: Construction of the optimum detection arc.

*C) Defining nasion intrinsic coordinates.*

To get the nasion intrinsic coordinates over the selected frame we apply the following equation, based on the geometric structure of the returned boxes by detectors, [Y-top, X-left, width, height], (see Fig.9), and the fact that we have the boxes structures for both eyes (Right eye [y1, x1, w1, h1], Left eye [y2, x2, w2, h2]) and the nose [y3, x3, w3, h3], then the nasion intrinsic coordinate is:

$$[x,y] = \left\{ \left[\frac{x2+x1}{2} + \frac{h2+h1}{4}\right], \left[\frac{y1+y2+w2}{2} + y3 + \frac{w3}{2}\right] \right\} \quad (1)$$

We also have the rotation angle depending on the selected best view.

The robustness of the facial detection consisted on the exploratory procedure by rotating around the three-dimensional object in one-degree angles. Only in the cases where the scanner object is inaccurate, the facial detection will fail.

*D) Pre-auricular points: Custom detectors*

To get the two dimensions intrinsic coordinates of the pre-auricular points, we need first to train a detector for each ear, as long as MATLAB® has several trained cascade classification models but not for ears. We use the Image Labeler app, from the Image Processing and Computer – Vision toolbox, to generate a Ground-Truth variable, and a "Vision Cascade Object Detector System" [21] to train the new detector we want.

The basics steps to achieve that goal are as follow:
1. To create a new session and define a new scene label.
2. Define a new region of interest (ROI), which is going to be the object we want to detect.
3. Add images to the session. These images need to contain the object of interest.
4. To manually select the ROI on each image, and apply to the current image.
5. Export Labels to the MATLAB® Workspace as a Ground-Truth variable.
6. Train the detector using the *trainCascadeObjectDtector* MATLAB® function.

   On this step we must: perform some transformation to the Ground-Truth variable; specify the full path to the positive samples used to define the ROI; specify the full path to negative samples, could be any image without the object to detect; to define the false alarm rate (FAR), and the number of cascade stages.

   In this work, besides the above input parameters, we used the "feature type" as 'HOG' (Histogram of Oriented Gradients), which is the default option of the *trainCascadeObjectDtector* MATLAB® function. This technique is based on the number of occurrences of gradient orientations in localized portions of an image [22]. It is computed on a dense grid of uniformity spaced cells and uses overlapping local contrast normalization for improved accuracy, achieving to find the features pattern of the ears as shown in Fig.6.

   During the training process, and before training each new stage, the function runs the detector consisting of the stages already trained on the negative images. Any objects detected from these images are false positives, which are used as negative samples. In this way, each new stage of the cascade is trained to correct mistakes made by previous stages [21].

   There are some important considerations when setting parameters to this function, to optimize the number of stages, the false positive rate and the true positive rate [21].

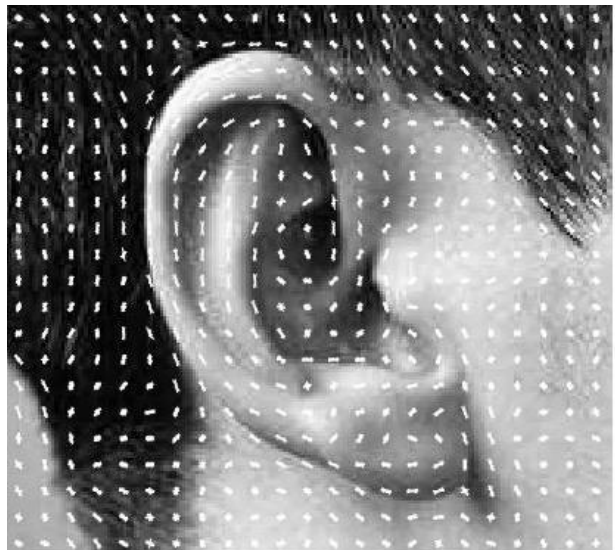

Fig.6: Histogram of Oriented Gradients of an ear.

There is a public script on GitHub to generate the detector once we have the Ground-Truth variable, and also a tutorial video of how to work on the Image Labeler app with the sessions, the scene labels and ROIs [23].

In the detectors, the stages are designed to reject negative samples as fast as possible. The assumptions is that the vast majority of the windows do not contain the object of interest. Conversely, true positive are rare but is necessary and worth to take the time to verify. The reasons are:
- A true positive occurs when a positive sample is correctly classified.
- A false positive occurs when a negative sample is mistakenly classified as positive.
- A false negative occurs when a positive sample is mistakenly classified as negative.

To work well, each stage in the cascade must have a low false negative rate. If a stage incorrectly labels an object as negative, the classification stops, and you cannot correct the mistake. However, each stage can have a high FAR. Even if the detector incorrectly labels a nonobject as positive, you can correct the mistake in subsequent stages.

The overall FAR of the cascade classifier is $f^s$, where $f$ is the FAR per stage in the range (0 1), and $s$ is the number of stages. Similarly, the overall true positive rate is $t^s$, where $t$ is the true positive rate per stage in the range (0 1). Thus, adding more stages reduces the overall FAR, but it also reduces the overall true positive rate. [21]

*E) Defining pre-auricular intrinsic coordinates.*

Now we apply those ear detectors to a projection of the surface rotated ± 87° from the rotation angle corresponding to the best view where the nasion was found. Once we detect ears as two dimension objects, taking into account that this detectors return boxes structures as [X-left Y-top width height], then we define the intrinsic coordinates of the "closest point" (*CP*) to the pre-auricular point as in equations (2) and (3) corresponding to the red point in Fig.7.

$$Right\_CP(x,y) = \{[X_{left} + width], [Y_{top} + \tfrac{height}{2}]\} \quad (2)$$

$$Left_{PP}\_CP(x,y) = \{[X_{left}], [Y_{top} + \tfrac{height}{2}]\} \quad (3)$$

Finally, given the detected closest point and considering the spatial location of the pre-auricular point shown in Fig.7 referred to the *CP*, we perform a correction according to equations (4) and (5) as a function of the box dimensions to get the best approximation of the real anatomical pre-auricular point.

$$Right\_PP(x,y) = \left\{ \begin{array}{l} [X_{left} + \tfrac{8}{10} * width], \\ [Y_{top} + 0.36 * height] \end{array} \right\} \quad (4)$$

$$Left\_PP(x,y) = \left\{ \begin{array}{l} [X_{left} + \tfrac{widht}{5}], \\ [Y_{top} + 0.36 * height] \end{array} \right\} \quad (5)$$

Equation (4) is defined taking into account the statistical results shown in Fig.8 corresponding to a set of one hundred different right ears and one hundred lefts ears. It is the variation of *x* and *y* when moving from the *CP* to the visual estimated pre-auricular point.

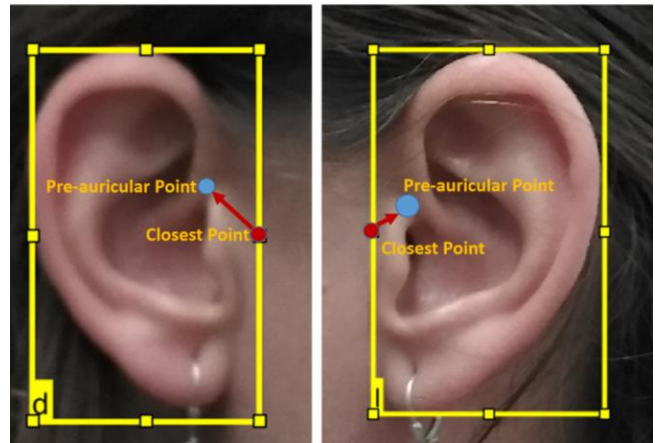

Fig.7: Correction of the pre-auricular points.

The correction consists of decreasing 0.2 of the width of the ear box and 0.14 of the height to the *x* and *y* coordinates of the CP respectively for the right ear, and increase 0.2 of the width and decrease 0.14 of the height of the detected box to the *x* and *y* coordinates respectively for the left ear.

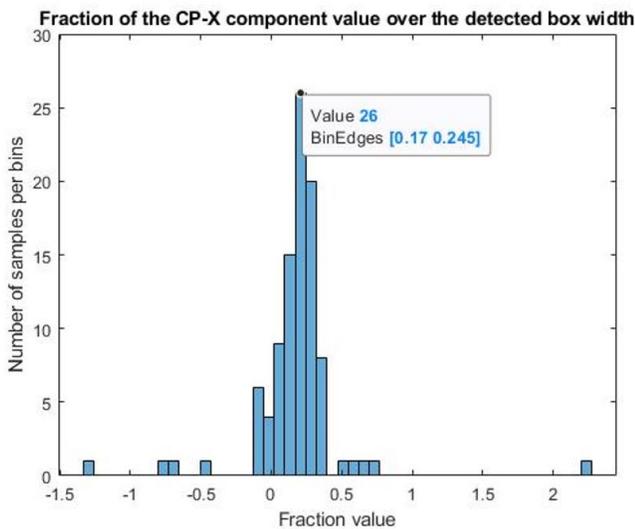
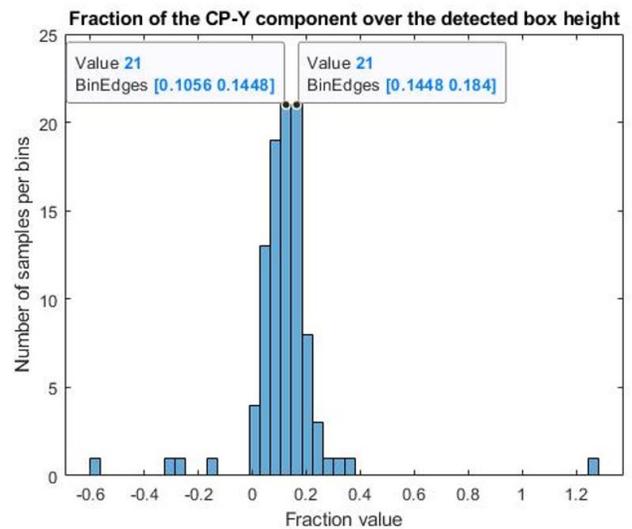

a)  b)

Fig.8: Histogram of the behavior of the distance between the pre-auricular point and the *CP* on the right ear. *a) x*-component variation and *b) y*-component variation.

This can be done as a consequence of the anatomical shape of the ears and the region of interest defined in the Image Labeler app during the training processes, and it is based on the fact that the 90% of the times the coordinates of the preauricular point are distant from the CP 0.14 of the detected box height in the *x* component of its coordinates, and 0.2 of the width of the boxes marked by detectors in the *y* component of its coordinate, as shown in Fig.8.

Since this is an estimation, an error will be incurred, but the estimated coordinates will always be more accurate than the *CP*, and we must say that fiducial land marks are not related to a specific coordinate because they are anatomical points that can be represented with several intrinsic coordinates corresponding to one pixel.

The performance of the calculation of the intrinsic coordinates of the pre-auricular points will be affected only by the occurrence of a false positive or false negative when detecting the ear, this probability will be analyzed in the in the process of designing the detectors at the results and discussion section. Nevertheless, the robustness of this procedure consists on the dynamical use of several ears detectors with different FARs and the recursive use of the combined set of possible detectors over a ten degrees arch of projected view, as will be mentioned later.

*F) Automatic re-projection*

At this point, we already have the fiducials points, but just a *(x,y)* point on an intrinsic coordinate system, regarding to a projection of the surface in two dimensions.
Now we need to control:

➢ The behavior of the three-dimensional object containing the head surface is controlled by rotation; once we know the angle of the view of the nasion, we define empirically as ± 87° the angles for the left and the right pre-auricular-point views respectively.

Additionally, we must ensure a 1:1 relation between the sizes of the figure pixels and the screen pixels, it is done with *truesize* MATLAB® function to the corresponding projection of the object view of interest, and applying the properties of the projection figure to the object figure after making the *truesize* to the figure.

➢ The mouse motion is controlled by changing the '*PointerLocation*' property of the Graphics Root Object (groot) of MATLAB® with the command (*set(groot,'PointerLocation',[x,y])*); note that we must do a correction between the coordinate system of the pixels on the screen and the coordinate system of the pixels on the image, this is done taking into account (Fig.9).

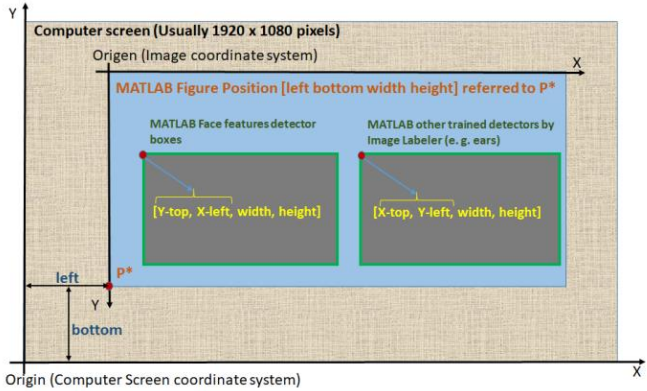

Fig.9: Differences of the intrinsic coordinate system of pixels on the screen, MATLAB® figures positions and boxes returned by detectors.

The mouse new position is corrected to be placed on the screen coordinate system and, it's done as function of the fiducial intrinsic coordinate in 2D, the axes position inside the figure, and the figure position inside the screen, according to the next relation:
Let:
The fiducial intrinsic coordinate be: $F = [F_x, F_y]$;
The figure position be:
$Fig = [Fig_{left}, Fig_{bottom}, Fig_{width}, Fig_{height}]$;
The current axes position be:
$Ax = [Ax_{left}, Ax_{bottom}, Ax_{width}, Ax_{height}]$;
And the number of rows on the images intrinsic coordinate system be $A$.
Then, the coordinates to send the mouse pointer will be:
$(x, y) = [F_y + Fig_{left} + Ax_{left},\ A - F_y + Fig_{bottom} + Ax_{bottom}]$ (6)

➢ The click event is performed by means of importing to MATLAB® the AWT (Abstract Window Toolkit), an API to develop GUI or Window-based applications in java [24].
Before programmatically clicking, we need to ensure of having the graphical object over all other windows.
Once we have captured the click event, next step is to find programmatically the (x,y,z) point selected over the surface. This is done with the FieldTrip private function *select3D* which is based on a projection transformation and the 2-D crossing test (Jordan Curve Theorem [25]).

*G) Create the coordinate system.*

Having the (*x,y,z*) coordinates corresponding to the nasion and both pre-auricular points, we generate the Subject Coordinate System (SCS/CTF), by means of the Fieldtrip function *ft_meshrealign,* which is based on the fiducials locations, as shown in Fig.10.

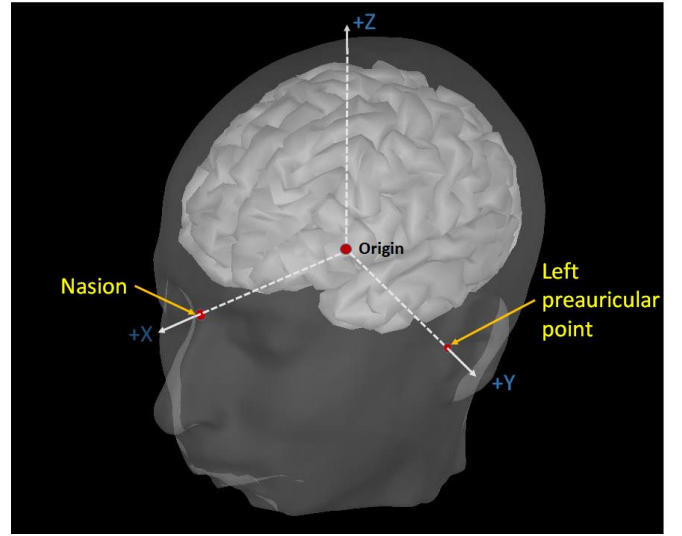

Fig.10: Subject coordinate system (SCS/CTF).

IV. RESULTS AND DISCUSSION

In this section we present the main results obtained on face features detection using the MATLAB own detectors, towards nasion definition, the results obtained on ear detection for preauricular points definition, both on 2D, the re-projection process and the generation of the coordinate system.

*A) Nasion location*

By definition, detectors always have a probability of not detecting an object (false negative) or detect an object for which it was not trained (false positive), as shown in Fig.11a). On this search procedure for facial features in the surface projection, the number of detected views depends only on the chosen rotation angle for turns step by step and the facial features conditions on the image pixels, Fig.11b).

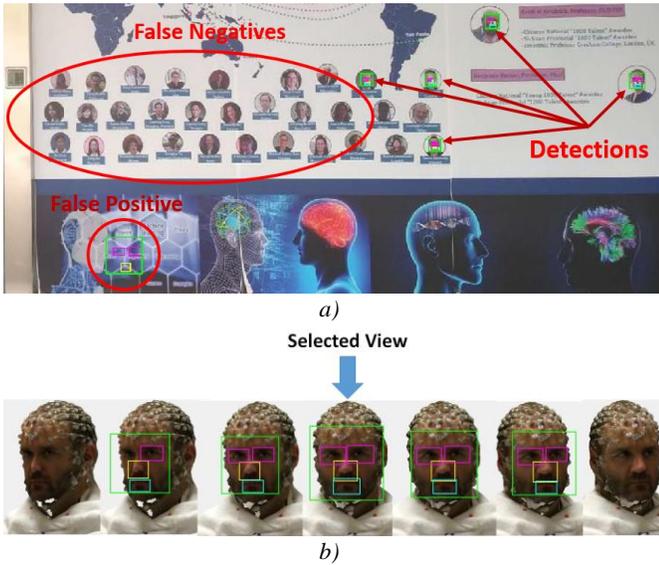

Fig.11: Face detection a) false positives and false negatives results for face features detection, b) best view selection.

The facial features were detected on the 100% of the analyzed surfaces when applying this detector to head views, with a minimum rotation range for consecutive detection of 3° referred to the frontal view, according to the detection efficacy in Table I, which is based on experimental results with nineteen subjects. Arch (°) represents the consecutive positions in which detection was successful.

Each subject will present a minimum rotation arch, depending on the facial features conditions on the surface of the object, but it does not mean that the detector is not able to find such characteristics outside that arch of the rotation angle. For instance, Fig.12 represents the dynamic of eye detection for subject number 16 on Table I, which represents the worst case among the subjects for view selection. Angles corresponding to arch from 148° to 171°, with values in zeros consecutively means that the left eye was not detected in that angles, and angles between 206° and 208°, with the same characteristics means that the right eye was not detected either; and this corresponds to the sighting of the left eye with the movement of rotation of the object, and the occultation of the left eye respectively.

TABLE I
FACE FEATURES DETECTION EFFICACY AS FUNCTION OF THE ROTATION ANGLE REFERRED TO THE FRONTAL VIEW

| Subject No. | Arch (°) | Selected angle (°) |
|---|---|---|
| 1 | 137 < α < 196 | 183 |
| 2 | 179 < α < 242 | 226 |
| 3 | 202 < α < 235 | 223 |
| 4 | 178 < α < 221 | 207 |
| 5 | 168 < α < 220 | 214 |
| 6 | 150 < α < 207 | 190 |
| 7 | 193 < α < 234 | 221 |
| 8 | 147 < α < 204 | 191 |
| 9 | 160 < α < 211 | 195 |
| 10 | 156 < α < 210 | 192 |
| 11 | 151 < α < 196 | 184 |
| 12 | 153 < α < 226 | 204 |
| 13 | 152 < α < 204 | 193 |
| 14 | 143 < α < 206 | 191 |
| 15 | 158 < α < 209 | 194 |
| 16 | **192 < α < 195** | **193** |
| 17 | 164 < α < 204 | 195 |
| 18 | 162 < α < 210 | 195 |
| 19 | 146 < α < 217 | 215 |

Given that the view corresponding to angle 193 turned out to be in which the eyes boxes represented the largest number of pixels within the image, for that reason the rank to which it belongs (192 < α < 195) is defined as the minimum for successful detection. This decision criteria is based on the fact that detected eyes on a view with an inclination angle always has less surface on the image, as shown in Fig.11b).

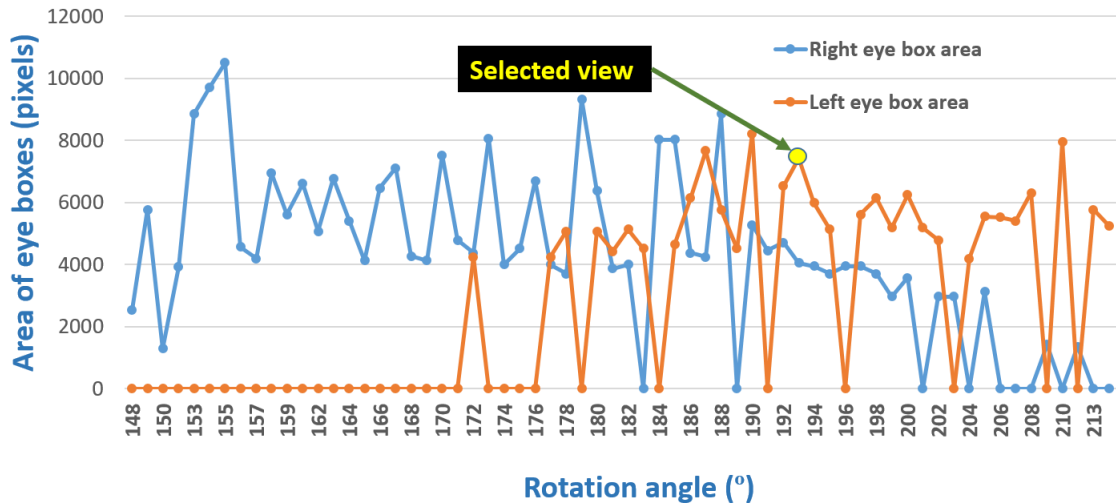

Fig.12: Dynamic of eye detection for subject 16 on Table II showing eyes boxes area vs. rotation angle.

To guarantee that we can find the facial features in a frontal view, we set the searching range between 0º and 360º.

Once we have the projection of the surface views, we need to choose the most symmetric according to the boxes of the detected feature, in such a way it allows to improve the accuracy of nasion location, as shown in Fig.11b), which is an example of the images sequence in which we found face features, and the selected one, based on the geometrical boxes symmetry.

In general, the face features detection success also depends on several elements when applyed to different kinds of images, such as: the face position, the presence of glasses, concealment of facial characteristics by excessive amount of hair, grimaces and other deformations. To avoid false positives, we recommend not to use objects around the subject head that could lead to pixelated images during the scanning process, such as a plaid shirt.

Finally, taking into account the nasion definition mentioned in previous section and applying equation (1), Fig.13 shows results of the face features detection and nasion coordinates according to the equation (1) as a function of detected boxes for the eyes and the nose while using a cascade object detector based on the classification models mentioned in the methods section, and having as decision criteria the boxes areas in pixels number and the symmetry. On the nasion coordinate we put a blank pixel highlighted by a red circle.

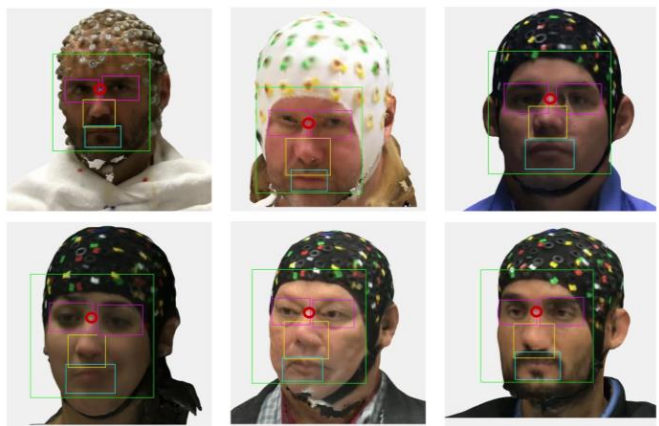

Fig.13: Face features detection and nasion location.

*B) Ears detection and definition of pre-auricular points*

Ears detectors were manually built as mentioned in the methods section. Using two sessions on the Image Labeler app in the Computer – Vision Toolbox, corresponding to 2308 positives images for the right ear, 2300 for the left ear, and 1323 negatives samples used for both detectors.

To design detectors with those images, the FAR was set to be: 0.3, 0.4, 0.5, 0.6 0.7, 0.8, 0.85 and 0.95 for each stages, as shown in Table II, in which also appear the number of stages obtained for each detector after training and the total FAR.

Taking into account these results, we propose to use all detectors dynamically, starting with the lowest FAR and switching to the next detector when no ear was detected, and to repeat this procedure recursively for a set of views corresponding to ten degrees range. This procedure will guarantee not to have false positives and a single truth positive.

TABLE II
DETECTORS DESIGN

| Type of detector | FAR per stage | Number of stages resulting | Total FAR |
|---|---|---|---|
| Left ear | 0.3 | 11 | 0.000002 |
| | 0.4 | 14 | 0.000003 |
| | 0.5 | 17 | 0.000008 |
| | 0.6 | 19 | 0.000061 |
| | 0.7 | 25 | 0.000134 |
| | 0.8 | 33 | 0.000634 |
| | **0.85** | **35** | **0.003386** |
| | 0.9 | 41 | 0.013303 |
| Right ear | 0.3 | 8 | 0.000066 |
| | 0.4 | 12 | 0.000017 |
| | 0.5 | 14 | 0.000061 |
| | 0.6 | 15 | 0.000470 |
| | 0.7 | 19 | 0.001139 |
| | **0.8** | **21** | **0.009223** |
| | 0.85 | 24 | 0.020233 |

Fig.14 shows the results obtained for both, right and left ears detection, as well as the CP in a yellow circle, and it´s correction towards the right pre-auricular point in a blue circle according to equations (4) and (5).

Another option to perform the pre-auricular point correction is to fit the detected box to the Fibonacci spiral, drawing at least four sections of the spirals and marking the fourth tangent point of the spiral.

This option was also used to make a visual verification of the obtained coordinates after applying equations (4) and (5), applying the φ proportion four consecutive times.

The fact that it is possible to apply the Fibonacci proportion to the detected box on the ears has another meaning; and it is that the learning procedure carried out by the detectors based on histograms of oriented gradients allows not only to detect invariant characteristics of the image but also the framing of the mentioned characteristics in a constant proportion.

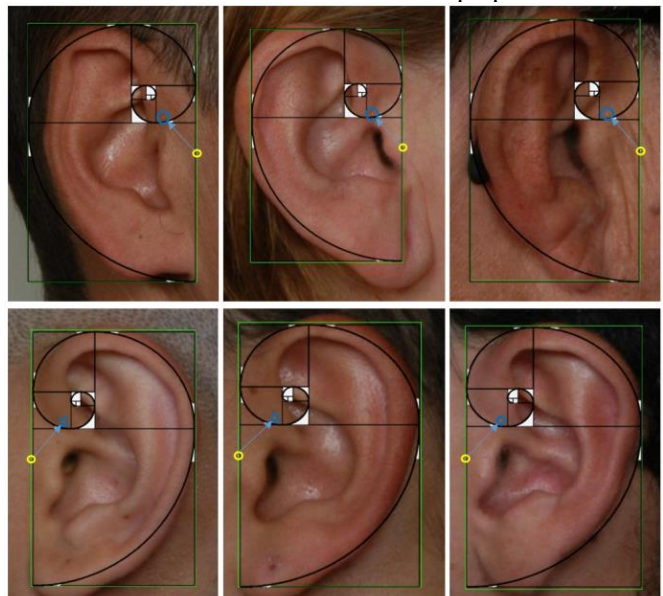

Fig.14: Ears detection results, correction of the CP toward the pre-auricular point and verification of the coordinates with the Fibonacci proportions.

### A) Re-projection process and Subject Coordinate System generation

To re-project 2D points to the 3D surface, corresponding to nasion and pre-auricular points, we implemented the procedure based on the camera model and the Jordan Curve Theorem and implemented by the FieldTrip toolbox functions as described in the methods section.

Finally, Fig.15 shows the SCS generated by means of the entire process of automatic fiducial detection described in this report.

With respect to the processing time required to define the SCS starting from the reading of the OBJ file, both the manual and the automatic procedures take an average of two minutes per subject on a Core-i7 personal computer with 16 Gigabytes of RAM.

The routines more time consume are those that read the object file and the exploratory procedure to detect facial features towards nasion coordinates.

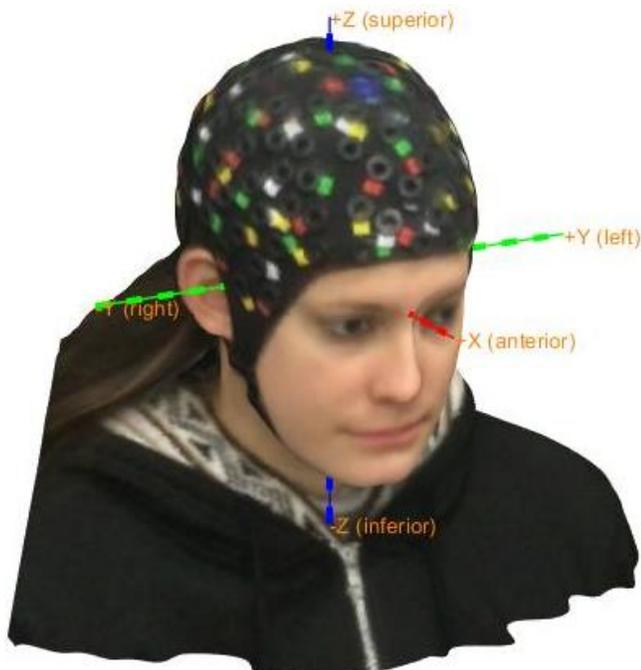

Fig.15: Subject Coordinate System.

We can say that the error incurred is due on the manual procedure precision since an anatomical point can be selected with multiple points belonging to an intrinsic coordinate system in an image.

### V. CONCLUSIONS

The fiducial location has been a problem for a long time when co-registering EEG with MRI. The method presented here has the advantages to be a cheap and fast alternative to create a reference system for spatial coordinates, and it will allow to define the EEG sensors coordinates based on real locations over the subjects head and not on templates.

Object detectors based on cascade classifiers offers a fast way to find facial features that allow to define fiducials land marks in two dimensions. The only limitation of this automatic fiducials location is related to the scanner procedure, because this must be done under good illumination conditions, making slow motion around the subject head to get all the facial features and guarantying that there are no artifacts that hide the facial characteristics of the subject, the detectors will do their job and the SCS will be generated successfully.

The robustness of detection algorithm proposed here consisted on the exploratory procedure for the facial features, and in the case of the ears it consisted on the dynamic use of the detectors beginning with those of lower FAR, switching to those with a higher FAR when detections do not occur and the recursively use of all detectors as function of the projection angle.

Since this is a proof of concept, the final SCS will allow us to search the electrode coordinates in a homogeneous reference coordinates system in future work. The same paradigm could be used to project the 3D object to 2D images with intrinsic coordinate system to find easily the electrodes and re-project to 3D generating the electrodes coordinates cloud.

### References

[1] L. Koessler, "Spatial localization of EEG electrodes". Clinical Neurophysiology, vol. 37. pp. 97-102, Apr. 2007.
[2] S. S. Dalal, "Consequences of EEG electrode position error on ultimate beamformer source reconstruction performance", Frontiers in Neurosciences, vol. 8. pp. 42, Mar. 2014.
[3] J. C. De Munck, "A practical method for determining electrode positions on the head". Electroencephalography and Clinical Neurophysiology, vol. 78. pp. 85-87, Jan. 1991.
[4] G. E. Chatrian, (2015, Feb.) Ten percent electrode system for topographic studies of spontaneous and evoked EEG activity. American Journal of EEG Technology. [Online]. 25(2), pp. 83-92. Available: https://www.tandfonline.com/doi/abs/10.1080/00029238.1985.11080163
[5] JG. H. Klem. "The ten twenty electrode system of the International Federation". Electroencephalography Clinical Neurophysiology", vol. 52. pp. 3-6. 1999.
[6] R. Oostenveld, "The five percent electrode system for high-resolution EEG and ERP measurements". Clinical Neurophysiology, vol. 112, pp. 713-719, Apr. 2001.
[7] R. Oostenveld, "FieldTrip: Open Source Software for Advanced Analysis of MEG, EEG, and Invasive Electrophysiological Data", Computational Intelligence and Neuroscience, vol. 2011.
[8] A. Delorme. (2004, Mar.). EEGLAB: an open source toolbox for analysis of single-trial EEG dynamics including independent component analysis, [Online]. Journal of Neuroscience Methods, 134(1), pp. 9-21. Available: https://www.sciencedirect.com/science/article/pii/S0165027003003479
[9] Magnetic Sensor System For Fast Response, High Resolution, High Accuracy, Three-Dimensional Position Measurements, by C. V. Nelson and B. C. Jacobs. (2004, Sep. 7). US Patent 6 789 043, [Online]. Available: https://patents.google.com/patent/US6789043
[10] © 2019 Occipital, Inc., "Structure Sensor, 3D Scanning. Augmented Reality. Instant Measurements". [Online]. Available: https://store.structure.io/buy/structure-sensor
[11] © 2018 Occipital, Inc. "What are the Structure Sensor's technical specifications?" [Online]. Available: https://support.structure.io/article/157-what-are-the-structure-sensors-technical-specifications
[12] R. Oostenveld, "Localizing electrodes using a 3D-scanner". [Online]. Available: http://www.fieldtriptoolbox.org/tutorial/electrode/
[13] "Mac & Apple Devices - EveryMac.com's Ultimate Mac Lookup". [Online]. Available: https://everymac.com/ultimate-mac-lookup/?search_keywords=A1893
[14] © Collins 2019, "Definition of 'nasion'". [Online]. Available: https://www.collinsdictionary.com/dictionary/english/nasion
[15] P. K. J. Yen, "Identification of Landmarks in Cephalometric Radiographs", The Angle Orthodontist, vol. 30, no. 1, pp. 35-41.
[16] Preauricular point | definition of preauricular point by Medical dictionary. [Online]. Available: https://medical-dictionary.thefreedictionary.com/preauricular+point
[17] Y. Miranda, "Detección de rostros empleando MATLAB". Revista SARANCE, vol. 39, pp. 41-49. Dec. 2017.

[18] ©1994-2019 The MathWorks, Inc. "Detect objects using the Viola-Jones algorithm", [Online]. Available: https://www.mathworks.com/help/vision/ref/vision.cascadeobjectdetector-system-object.html
[19] Y. Freund. (1997, Aug.). "A decision-theoretic generalization of on-line learning and an application to boosting". Journal of Computer and System Sciences. [Online]. 55(1), pp. 119-139. Available: https://www.sciencedirect.com/science/article/pii/S002200009791504X
[20] P. Viola and M. J. Jones, "Rapid Object Detection using a Boosted Cascade of Simple Features", in *Proceedings of the 2001 IEEE Computer Society Conference on Computer Vision and Pattern Recognition*, 2001, pp. 511-518.
[21] ©1994-2019. The MathWorks, Inc., "Train a Cascade Object Detector", [Online]. Available: https://la.mathworks.com/help/vision/ug/train-a-cascade-object-detector.html
[22] N. Dalal, and B. Triggs, "Histograms of Oriented Gradients for human detection." in *IEEE Comp. Soc. Conf. on Comp. Vision and Patt. Recog.* 2005, San Diego, US, pp. 886-893.
[23] "Matlab code for automatic fiducial detection with computer vision", E. E. Gallego. [Online]. Available: https://github.com/elieserernesto/Fiducial_Detection
[24] "Java Abstract Window Toolkit Tutorial". [Online], Available: https://www.javatpoint.com/java-awt
[25] C. Jordan, "Jordan's Cours D'Analyce" in "Cours D'Analyse de L'˙Ecole Polytechnique", 2nd edition, Paris, France, 189.